\documentclass[pra,showpacs,twocolumn]{revtex4}

\usepackage{auto-pst-pdf}
\usepackage{amsmath}
\usepackage{amssymb,amscd}
\usepackage{psfrag}
\usepackage{graphicx}
\usepackage{braket}
\usepackage[dvips]{epsfig}
\usepackage{epsfig}

\usepackage{subfigure}
\usepackage{color}

\begin{document}
\title{Continuous variable measurement device independent multipartite quantum communication}
\author{Yadong Wu$^1$, Jian Zhou$^2$, Xinbao Gong$^1$, Ying Guo$^2$, Zhi-Ming Zhang$^3$, Guangqiang He$^{1,3,4}$}
\email[Email: ]{gqhe@sjtu.edu.cn}
 \affiliation{$^1$State Key Laboratory of Advanced Optical Communication Systems  and Networks, Department of Electronic Engineering, Shanghai Jiao Tong University, Shanghai {\rm 200240}, PR China,  \\
 $^2$School of Information Science and Engineering,
Central South University, Changsha {\rm 410083}, PR China, \\
 $^3$Guangdong Provincial Key Laboratory of Quantum Engineering and Quantum Materials,   South China Normal University, Guangzhou {\rm 510006}, PR China, \\
 $^4$State Key Laboratory of Precision Spectroscopy, East China Normal University, Shanghai {\rm 200062}, PR China}

\begin{abstract}
A continuous variable measurement device independent multi-party quantum communication protocol is investigated in this paper. Utilizing distributed continuous variable Greenberger-Horne-Zeilinger state, this protocol can implement both quantum cryptographic conference and quantum secret sharing. We analyze the security of the protocol against both entangling cloner attack and coherent attack. Entangling cloner attack is a practical individual attack, and coherent attack is the optimal attack Eve can implement. Simulation results show that coherent attack can greatly reduce the secret key rate. Different kinds of entangled attacks are compared and we finally discuss the optimal coherent attacks.
\end{abstract}
\pacs{03.67.Dd, 03.67.Hk}
\maketitle

\section{Introduction}
In quantum cryptography, to maximize secure transmission distance and remove detector side attacks, physicists use  measurement device independent (MDI) method~\cite{PhysRevLett.108.130503}, which has been experimentally realized \cite{PhysRevLett.111.130502}.
In MDI quantum key distribution (QKD), anyone participating the quantum communication connects to an untrusted party, who is not a legitimate member in the quantum communication. The secure communication relies on the untrusted party's measurement.
So attacks on measurement devices are moved from legitimate members' sides to the untrusted party's side. Since any attack on measurement devices can be transformed into some attack in the channel followed by a correctly operated measurement~\cite{pirandola2015high}, we can just consider attacks in the channels. One important realistic attack is entangling cloner attack, which utilizes EPR state to maximize the information Eve can steal in individual attack~\cite{Grosshans}. The optimal attack Eve can implement, however, is not individual attack, but coherent attack, where Eve uses the ancillary system to globally interact with the signals and finally makes an optimal joint measurement.

MDI multipartite quantum communication with long distance is investigated in
ref.~\cite{PhysRevLett.114.090501}. This research is based on discrete variable systems, while Gaussian modulation and homodyne measurement~\cite{RevModPhys.84.621}
provide us another way to realize MDI multipartite quantum communication in continuous variable (CV) quantum systems. In this paper, we use CV to investigate multi-party quantum cryptography. CV MDI two-party quantum cryptography has been investigated in ref.~\cite{pirandola2015high}. Instead of using coherent state and heterodyne measurement as ref.~\cite{pirandola2015high}, our protocol utilizes squeezed state of light and homodyne measurement to maximize the secret key rate. Hence the main difficulty for the practical realization of our protocol is to generate squeezed state of light. Although it is much more difficult than generating coherent state of light, some experiments on CV squeezed states have been done~\cite{madsen2012continuous,PhysRevLett.113.060502}, indicating that CV quantum communication based on squeezed state can be realized in future.

We design and investigate two kinds of CV MDI multipartite quantum communication protocols in this paper. One is  quantum cryptographic conference (QCC)~\cite{PhysRevA.57.822} and the other is quantum secret sharing (QSS)~\cite{PhysRevA.59.1829}. QCC enables each individual within a specific group to decrypt the encrypted messages published by any group member, whereas nobody outside the group can successfully decrypt the secret messages.  QSS enables an authorized group of people to decrypt the secret messages by collaboration, but any unauthorized group of people fails to decrypt the messages.

This paper is organized as follows. Section \ref{sec1} introduces the MDI multipartite quantum communication protocols in detail. Section \ref{sec2} analyzes the security of this protocol against entangling cloner attack and coherent attack respectively for both QCC and QSS. Section \ref{sec3} shows the numerical simulation of this protocol against two kinds of attacks. Section \ref{sec4} gives the conclusion of this paper.

\section{Protocols of multipartite quantum communication}
\label{sec1}
We are going to explain the details of the protocols for both QCC and QSS in this section. Both of them rely on the post-processed GHZ state, while the main difference is within the postprocessing of classical data.

\begin{figure}[htbp]
  \begin{minipage}[t]{0.4\textwidth}
  \centering
  \includegraphics[width=0.9\textwidth]{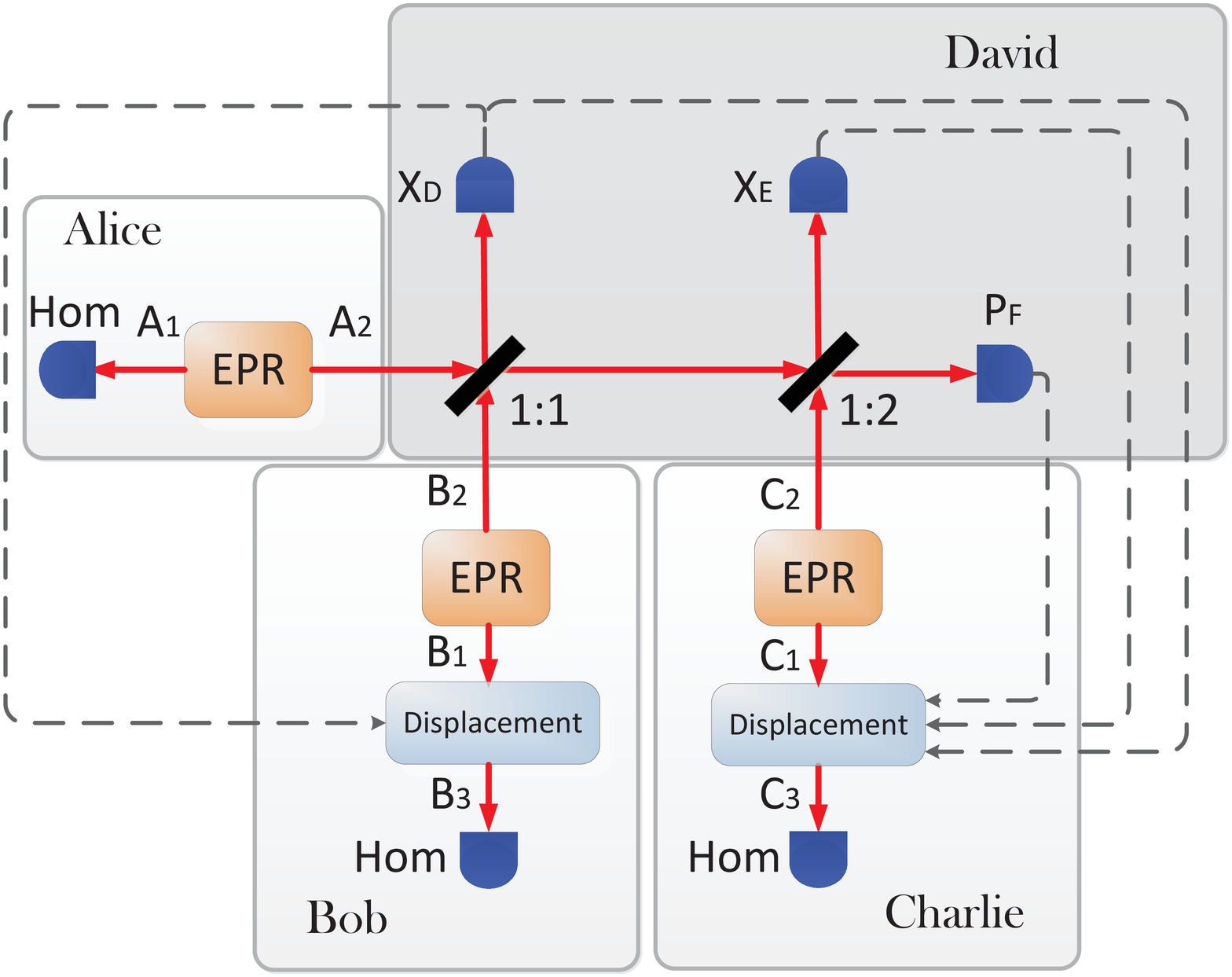}\\
  \caption{(Color online) EB scheme }\label{fig1}
\end{minipage}
\begin{minipage}[t]{0.4\textwidth}
\centering
  \includegraphics[width=0.9\textwidth]{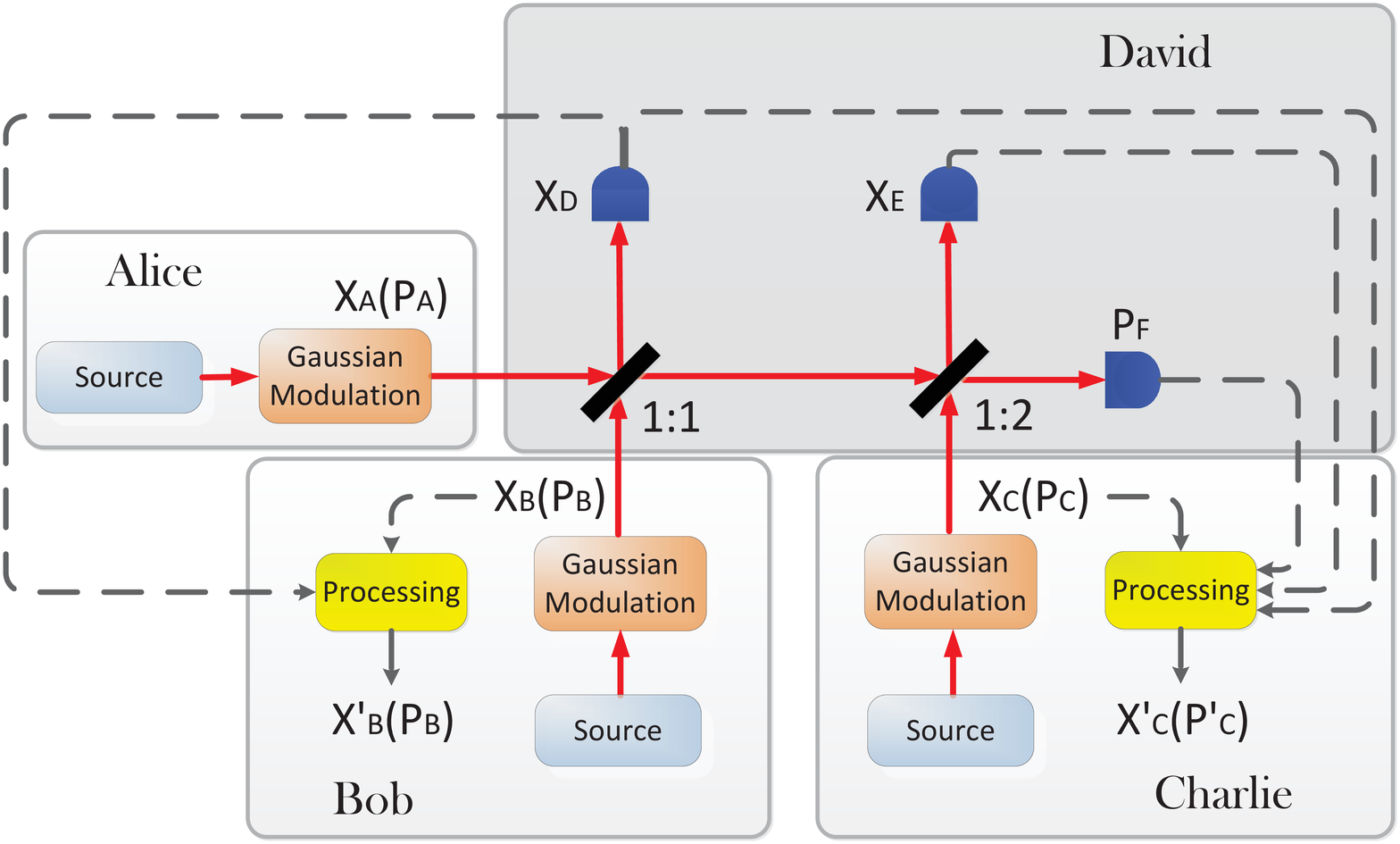}\\
  \caption{(Color online) PM scheme }\label{fig2}
\end{minipage}
\end{figure}

Before going into the detail of the protocol, we want to first introduce CV Greenberger-Horne-Zeilinger (GHZ) state~\cite{greenberger1989going}.
To implement these two kinds of multi-party quantum communication protocols, we utilize CV GHZ state, which is theoretically investigated~\cite{PhysRevA.67.052315} and experimentally realized by linear optics~\cite{PhysRevLett.90.167903,PhysRevLett.91.080404}.
It is a multipartite entangled state whose uncertainties of relative position  and total momentum are squeezed. For tripartite CV GHZ state, their positions and momenta satisfy the relations: $\hat{X}_1-\hat{X}_2\rightarrow 0,\hat{X}_2-\hat{X}_3\rightarrow 0$ and $\hat{P}_1+\hat{P}_2+\hat{P}_3\rightarrow 0$. CV GHZ state can be generated by a series of beam splitters with particular transmittances and squeezed vacuum states~\cite{PhysRevA.73.032318}. But in our protocol, CV GHZ state is not prepared and distributed, whereas is obtained by postprocessing using the idea of entanglement swapping~\cite{PhysRevA.61.010302,PhysRevLett.83.2095}. The whole protocol of quantum commincation is shown in Fig.~\ref{fig1}.

Two schemes of the protocol are shown in the following. One is entanglement-based (EB) scheme and the other is prepare-and-measurement (PM) scheme. In both EB and PM schemes, Alice, Bob and Charlie are connected with a fourth, untrusted person David and the secure communication relies on David's measurements. It implies that both schemes are MDI multipartite quantum communication protocols, which remove any detector side attack in Alice's, Bob's and Charlie's sides. We introduce the EB scheme first,  then the PM scheme.

Alice, Bob and Charlie prepare an EPR pair respectively, which, in CV case, is a two-mode squeezed state
(TMSS)~\cite{PhysRevLett.59.2555}.  They hold one mode of the EPR pair in their own possession and send the other mode to the fourth, untrusted person David. Receiving three modes from Alice, Bob and Charlie, David does the following operation. The two modes from Alice and Bob go through a beam splitter with transmittance $\frac{1}{\sqrt{2}}$. Then the position quadrature $\hat{X}_D$ of the output mode $D$ is detected by a homodyne measurement. The other output mode is mixed with the mode from Charlie by another beam splitter with transmittance $\sqrt{\frac{2}{3}}$ and the output modes  are denoted by $E$ and $F$. The position quadrature $\hat{X}_E$ and the momentum quadrature $\hat{P}_F$ are detected by two homodyne measurements. David publishes the measurement outcomes $X_D, X_E$ and $P_F$. To construct a  GHZ state, Bob and Charlie use these data to finish the displacement operations on their own modes. Specifically, Bob shifts the position quadrature  $\hat{X}_{B_1}$ with $\sqrt{2}X_D$ and Charlie shifts the position $\hat{X}_{C_1}$  and the momentum $\hat{P}_{C_1}$   with $\left(\sqrt{\frac{1}{2}}X_D-\sqrt{\frac{3}{2}}X_E\right)$ and $\sqrt{3}P_F$ respectively. After that, Alice, Bob and Charlie own  the modes $A_1, B_3$ and $C_3$ respectively and these three modes form a distributed CV GHZ state. As for QCC scheme,  they apply homodyne measurements over the positions respectively, and use the measurement outcomes $X_{A_1}, X_{B_3}$ and $X_{C_3}$ to do reconciliation and postselection. Since $\hat{X}_{A_1}-\hat{X}_{B_3}\rightarrow 0$ and $\hat{X}_{B_3}-\hat{X}_{C_3}\rightarrow 0$, they can obtain coincident keys. As for QSS scheme, they homodyne the momentum quadratures respectively. At least two of the three must share their measurement outcomes and do reconciliation and postselection with the third person. Because of the relation $\hat{P}_{A_1}+\hat{P}_{B_3}+\hat{P}_{C_3}\rightarrow 0$, they can obtain the third person's secret key. In both QCC and QSS schemes, with the random secret keys, they can use one-time pad~\cite{shannon1949communication}
to implement unconditional secure multipartite quantum communication.

Now we introduce the equivalent PM scheme.  Alice, Bob and Charlie firstly  generate Gaussian-distributed random numbers respectively and keep these data private. In QCC, Alice, Bob and Charlie do Gaussian modulation on position-squeezed vacuum states so that the mean positions of the modulated squeezed states become Gaussian-distributed random numbers $X_A, X_B$ and $X_C$. In QSS, they do Gaussian modulation on momentum-squeezed vacuum states so that the mean momenta of the modulated squeezed states become Gaussian-distributed random numbers $P_A, P_B$ and $P_C$. This preparation of a Gaussian modulated squeezed state is equivalent to making single-mode homodyne measurement over a TMSS. This is because
as for a TMSS, single-mode homodyne detection projects the other mode into a squeezed state with a specific mean value related to the measurement outcome.
Then the Gaussian modulated squeezed states are sent to the fourth, untrusted person David. David mixes these three modes by two specific beam splitters, makes homodyne measurements on the three outputs and publishes the measurement outcomes, same as the EB scheme. Bob and Charlie use the  public data to postprocess their own data.  In QCC, Alice remains her data constant, Bob modifies $X_B$ as $X'_B=X_B+\sqrt{2}X_D$ and Charlie modifies $X_C$ as $X'_C=X_C+\left(\sqrt{\frac{1}{2}}X_D-\sqrt{\frac{3}{2}}X_E\right)$, so that their data satisfy the relation  $X_A-X'_B\rightarrow 0$ and $X'_B-X'_C\rightarrow 0$. By doing reconciliation and postprocessing, they can obtain the coincident keys for QCC.  In QSS, Alice and Bob remain their data unchanged and Charlie replaces $P_C$ with $P'_C=P_C+\sqrt{3}P_F$,  making their data satisfy $P_A+P_B+P'_C\rightarrow 0$. Finally, two of them share their private data with each other. By reconciliation and postprocessing, they can obtain the secret key of the third person.

\section{Security analysis}
\label{sec2}
Now let's investigate the security of this protocol for both QCC and QSS schemes. The security of EB and PM schemes are equivalent. Since in EB scheme, we can use the method of purification to calculate the secret key rate, we choose to analyze the EB scheme. Our security analysis involves two kinds of attacks: one is entangling cloner attack and the other is coherent attack.

In general, there can be any attack at the detector side, i.e. at David's side in Fig.~\ref{fig1}. But any attack at the detector side is equivalent to adding a specific attack in the channel followed by a correctly operated measurement device~\cite{pirandola2015high}. Thus we can assume that Eve's attack only exists in the channel and that David's operation and measurement data can be trusted.

\subsection{Independent entangling cloner attack}
\begin{figure}[htbp]
  \includegraphics[width=0.4\textwidth]{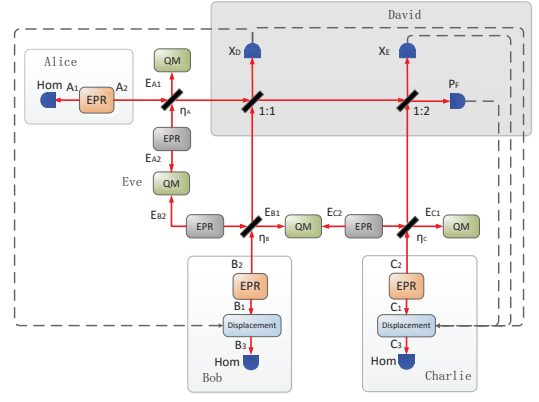}\\
  \caption{(Color online) EB scheme against independent entangling cloner attacks}\label{fig3}
\end{figure}

In this subsection, we focus on independent entangling cloner attack in each channel.  As shown in Fig.~\ref{fig3}, at the beginning, Eve owns three independent EPR pairs, i.e., three TMSS's. He injects one mode of each pair into the channel, through a beam splitter with transmittance $\eta_A (\eta_B, \eta_C)$, and stores the output mode of each beam splitter, $\hat{E}_{A1\,(B1, C1)}$, and the other mode of each EPR pair, $\hat{E}_{A2\,(B2, C2)}$ , in the quantum memory.

In QCC, we consider the case when Alice wants to send her secret message to Bob and Charlie. To achieve this, Alice needs to share secret keys with Bob and Charlie, respectively, by implementing QKD.
The secret key rate can be defined as
\begin{equation} \label{eq1}
K_{QCC}=\min\{K_{AB},K_{AC}\},
\end{equation} where
$K_{AB}$ is the secret key rate between Alice and Bob, and $K_{AC}$
is the secret key rate between Alice and Charlie.

Two kinds of reconciliation methods lead to different secret key rates.
With reverse reconciliation,
\begin{equation}\label{eq3}
\begin{aligned}
K_{AB}^{RR}&=\beta I(X_{A_1}:X_{B_3})- I(X_{A_1}:X_{E_{A1}},X_{E_{A2}}),\\
K_{AC}^{RR}&=\beta I(X_{A_1}:X_{C_3})- I(X_{A_1}:X_{E_{A1}},X_{E_{A2}}).
\end{aligned}
\end{equation}
With direct reconciliation,
\begin{equation}\label{eq4}
\begin{aligned}
K_{AB}^{DR}&=\beta I(X_{A_1}:X_{B_3})- I(X_{B_3}:X_{E_{B1}},X_{E_{B2}}),\\
K_{AC}^{DR}&=\beta I(X_{A_1}:X_{C_3})- I(X_{C_3}:X_{E_{C1}},X_{E_{C2}}).
\end{aligned}
\end{equation}
At the righthand sides of Eqs.~(\ref{eq3}) and Eqs.~(\ref{eq4}), the first term  represents the mutual information between the measurement data of $X_{A_1}$ and the measurement data of $X_{B_3(C_3)}$~\cite{garcia2007quantum}, and the second term denotes the mutual information between the measurement data of $X_{A_1 (B_3, C_3)}$ and the measurement data of $X_{E_{A1(B1,C1)}}$ and $X_{E_{A2(B2,C2)}}$. $\beta$ is the reconciliation efficiency.

In QSS,  we assume Charlie holds the secret key, and Alice and Bob have to collaborate with each other to obtain the secret key. The secret key rate can be defined as
\begin{equation}\label{eq5}
\begin{aligned}
K_{QSS}^{RR}
=\beta I(P_{A_1},P_{B_3}:P_{C_3})-I(P_{C_3}:P_{E_{C1}},P_{E_{C2}}).
\end{aligned}
\end{equation}
with reverse reconciliation, and
\begin{equation}\label{eq10}
\begin{aligned}
K_{QSS}^{DR}
=&\beta I(P_{A_1},P_{B_3}:P_{C_3})-I(P_{A_1}:P_{E_{A1}},P_{E_{A2}})\\
-&I(P_{B_3}:P_{E_{B1}},P_{E_{B2}}).
\end{aligned}
\end{equation}
with direct reconciliation. At the righthand sides of Eqs.~(\ref{eq5}) and Eqs.~(\ref{eq10}), the first term  represents the mutual information between the measurement data of $P_{A_1}$ and the measurement data of $P_{B_3}$ and $P_{C_3}$, and the second term denotes the mutual information between the measurement data of $P_{A_1 (B_3, C_3)}$ and the measurement data of $P_{E_{A1(B1,C1)}}$ and $P_{E_{A2(B2,C2)}}$.

To calculate the mutual information in Eqs. (\ref{eq3}), (\ref{eq4}), (\ref{eq5}) and (\ref{eq10}), we need to obtain the covariance matrix of the whole state held by Alice, Bob, Charlie and Eve in the following way.

At the beginning of this protocol, the initial whole state $\mathbf{\rho}_{A,E_A,B,E_B,C,E_C}$ is the tensor product of six independent TMSS's. Its covariance matrix is
\begin{equation}
 \mathbf{V}_{A,E_A,B,E_B,C,E_C}=\bigoplus_{k=1}^3\mathbf{V},
\end{equation}
where
\begin{equation}
\mathbf{V}=
\left(
  \begin{array}{cccc}
    V\mathbf{I} & \sqrt{V^2-1}\mathbf{Z} & \mathbf{0} & \mathbf{0} \\
    \sqrt{V^2-1}\mathbf{Z} & V\mathbf{I} & \mathbf{0} & \mathbf{0} \\
    \mathbf{0} & \mathbf{0} & V_E\mathbf{I} & \sqrt{V_E^2-1}\mathbf{Z} \\
    \mathbf{0} & \mathbf{0} & \sqrt{V_E^2-1}\mathbf{Z} & V_E\mathbf{I} \\
  \end{array}
\right).
\end{equation}
$V\,(V\ge 1)$ is the variance of Alice's (Bob's, Charlie's) TMSS's, $V_E\,(V_E\ge 1)$ is the variance of Eve's TMSS's,
$\mathbf{I}$ is  identity matrix, $\mathbf{0}$ is zero matrix and $\mathbf{Z}$ is the Pauli Z matrix.

In each channel, Alice's (Bob's, Charlie's) mode goes through a beam splitter with transmittance $\eta_A(\eta_B, \eta_C)$.
The overall operation of these three beam splitters can be written as
\begin{equation}\label{eq12}
\mathbf{U_{Eve}}=
\mathbf{BS_A}\bigoplus \mathbf{BS_B}\bigoplus \mathbf{BS_C},
\end{equation}
where
\begin{equation}\label{eq13}
\mathbf{BS_{A(B,C)}}=
\left(
  \begin{array}{cccc}
    \mathbf{I} & \mathbf{0} & \mathbf{0} & \mathbf{0} \\
    \mathbf{0} & \sqrt{\eta_{A(B,C)}}\mathbf{I} & \sqrt{1-\eta_{A(B,C)}}\mathbf{I} & \mathbf{0} \\
    \mathbf{0} & -\sqrt{1-\eta_{A(B,C)}}\mathbf{I} & \sqrt{\eta_{A(B,C)}}\mathbf{I} & \mathbf{0} \\
    \mathbf{0} & \mathbf{0} & \mathbf{0} & \mathbf{I} \\
  \end{array}
\right).
\end{equation}

At David's side,  the overall operation of the two beam splitters is
 \begin{equation}\label{eq14}
\mathbf{U_{David}}=\mathbf{BS_2}\mathbf{BS_1},
\end{equation}
where
\begin{equation}\label{eq15}
\begin{aligned}
\mathbf{BS_1}&=
\left(
  \begin{array}{ccc}
    \mathbf{W_1} & \mathbf{W_2} & \mathbf{0} \\
    -\mathbf{W_2} & \mathbf{W_1} & \mathbf{0} \\
    \mathbf{0} & \mathbf{0} & \mathbf{I} \\
  \end{array}
\right),
\mathbf{BS_2}=
\left(
  \begin{array}{ccc}
    \mathbf{W_3} & \mathbf{0} & \mathbf{W_4} \\
    \mathbf{0} & \mathbf{I} & \mathbf{0} \\
    -\mathbf{W_4} & \mathbf{0} & \mathbf{W_3} \\
  \end{array}
\right),\\
\mathbf{W_1}&=
\left(
  \begin{array}{cccc}
    \mathbf{I} & \mathbf{0} & \mathbf{0} & \mathbf{0} \\
    \mathbf{0} & \mathbf{\frac{1}{\sqrt{2}}I} & \mathbf{0} & \mathbf{0} \\
    \mathbf{0} & \mathbf{0} & \mathbf{I} & \mathbf{0} \\
    \mathbf{0} & \mathbf{0} & \mathbf{0} & \mathbf{I} \\
  \end{array}
\right),
\mathbf{W_2}=
\left(
  \begin{array}{cccc}
    \mathbf{0} & \mathbf{0} & \mathbf{0} & \mathbf{0} \\
    \mathbf{0} & \mathbf{\frac{1}{\sqrt{2}}I} & \mathbf{0} & \mathbf{0} \\
    \mathbf{0} & \mathbf{0} & \mathbf{0} & \mathbf{0} \\
    \mathbf{0} & \mathbf{0} & \mathbf{0} & \mathbf{0} \\
  \end{array}
\right),\\
\mathbf{W_3}&=
\left(
  \begin{array}{cccc}
    \mathbf{I} & \mathbf{0} & \mathbf{0} & \mathbf{0} \\
    \mathbf{0} & \mathbf{\sqrt{\frac{2}{3}}I} & \mathbf{0} & \mathbf{0} \\
    \mathbf{0} & \mathbf{0} & \mathbf{I} & \mathbf{0} \\
    \mathbf{0} & \mathbf{0} & \mathbf{0} & \mathbf{I} \\
  \end{array}
\right),
\mathbf{W_4}=
\left(
  \begin{array}{cccc}
    \mathbf{0} & \mathbf{0} & \mathbf{0} & \mathbf{0} \\
    \mathbf{0} & \mathbf{\frac{1}{\sqrt{3}}I} & \mathbf{0} & \mathbf{0} \\
    \mathbf{0} & \mathbf{0} & \mathbf{0} & \mathbf{0} \\
    \mathbf{0} & \mathbf{0} & \mathbf{0} & \mathbf{0} \\
  \end{array}
\right).
\end{aligned}
\end{equation}

Before any homodyne measurement, the whole state becomes $\mathbf{\rho}_{{A_1,F,E_{A1},E_{A2},B_1,D,E_{B1},E_{B2},C_1,E,E_{C1},E_{C2}}}$ with covariance matrix
\begin{equation}\label{eq7}
\begin{aligned}
&\mathbf{V}_{A_1,F,E_{A1},E_{A2},B_1,D,E_{B1},E_{B2},C_1,E,E_{C1},E_{C2}}\\
=&\mathbf{U}_{David}\mathbf{U}_{Eve} \mathbf{V}_{A,E_A,B,E_B,C,E_C}\mathbf{U}_{Eve}^T\mathbf{U}_{David}^T.
\end{aligned}
\end{equation}

By permutating the modes in the covariance matrix in Eq.~(\ref{eq7}) in the order of $A_1,B_1, C_1, Eve, D, E, F$, we can rewrite it in the form of
\begin{equation}
\begin{aligned}
&\mathbf{V}_{A_1,B_1, C_1,Eve,D,E,F}=\\
                             &\left(
                               \begin{array}{cc}
                               \mathbf{V}_{A_1,B_1,C_1,Eve,D,E} &\mathbf{C} \\
                               \mathbf{C}^T & \mathbf{V}_F \\
                               \end{array}
                             \right),
\end{aligned}
\end{equation}
where the subscript $Eve$ denotes all the six modes $E_{A1}, E_{A2}, E_{B1}, E_{B2}, E_{C1}$ and $E_{C2}$, and $\mathbf{C}$  represents covariance submatrix.

Homodying $\hat{P}_F$ turns the reduced covariance matrix $\mathbf{V}_{A_1,B_1,C_1,Eve,D,E}$ into~\cite{PhysRevLett.89.137903}
\begin{equation}\label{eq8}
\begin{aligned}
 &\mathbf{V}_{A_1,B_1,C_1,Eve,D,E|P_F}\\
  =&
  \mathbf{V}_{A_1,B_1,C_1,Eve,D,E}
  -\mathbf{C}\left(
               \begin{array}{cc}
                 0 & 0 \\
                 0 & \frac{1}{V(\hat{P}_F)} \\
               \end{array}
             \right)\mathbf{C}^T,
\end{aligned}
\end{equation}
where $V(\hat{P}_F)$ is the variance of $\hat{P}_F$, given in the matrix $\mathbf{V}_F$. As shown in Eq.(\ref{eq8}), the covariance matrix following partial homodyne measurement has nothing to do with the measurement outcome. Thus, although the measurement result may be different each time, the covariance matrix following the partial measurement remains the same.

Iteratively calculating the covariance matrix of the state after partial Gaussian measurements~\cite{RevModPhys.84.621,Covariance}, we can obtain the covariance matrix of the partial state $\mathbf{\rho}_{A_1,B_1,C_1,Eve}$ after $\hat{X}_D, \hat{X}_E$ and $\hat{P}_F$ are homodyned.

Since displacement operations $e^{i\xi\hat{X}}$ and $e^{i\xi'\hat{P}}$ remain the variances and covariances of $\hat{X}$ and $\hat{P}$ the same, while only changes their mean values, the partial state $\mathbf{\rho}_{A_1,B_3,C_3}$ owns the same covariance matrix as $\mathbf{\rho}_{A_1,B_1,C_1|X_D,X_E,P_F}$. Thus,
by now, we have obtained the covariance matrix of the  state $\mathbf{\rho}_{A_1,B_3,C_3,Eve}$, denoted by $\mathbf{V}_{A_1,B_3,C_3,Eve}$.

Now we can calculate Eqs. (\ref{eq3}), (\ref{eq4}), (\ref{eq5}) and (\ref{eq10}) by using $\mathbf{V}_{A_1,B_3,C_3,Eve}$. In this calculation, we may need to obtain the covariance matrix of a reduced state of $\mathbf{\rho}_{A_1,B_3,C_3,Eve}$ by using the formula similar to Eq. (\ref{eq8}), when a partial Homodyne measurement is applied.

The first terms at the righthand sides of Eqs. (\ref{eq3}) and (\ref{eq4}) are given by
\begin{equation}\label{eq11}
I(X_{A_1}:X_{B_3(C_3)})=\frac{1}{2}\log_2{\frac{V(\hat{X}_{B_3(C_3)})}{V(\hat{X}_{B_3(C_3)}|X_{A_1})}},
\end{equation}
where $V(\hat{X}_{B_3(C_3)}|X_{A_1})$ is the conditional variance of $\hat{X}_{B_3(C_3)}$ after $\hat{X}_{A_1}$ is homodyned and can be obtained from the covariance matrix $\mathbf{V}_{B_3C_3|X_{A_1}}$.

The second terms in Eqs.~(\ref{eq3}) and Eqs.~(\ref{eq4}) are
\begin{equation}\label{eq9}
\begin{aligned}
&I(X_{A_1(B_3, C_3)} :X_{E_{A1(B1,C1)}},X_{E_{A2(B2,C2)}})\\
=&\frac{1}{2}\log_2{\frac{V(\hat{X}_{A1(B1,C1)})}{V(\hat{X}_{A1(B1,C1)}|X_{E_{A1(B1,C1)}}, X_{E_{A2(B2,C2)}})}},
\end{aligned}
\end{equation}
where $V(\hat{X}_{A1(B1,C1)}|X_{E_{A1(B1,C1)}}, X_{E_{A2(B2,C2)}})$ is the variance of $\hat{X}_{A1(B1,C1)}$ after $\hat{X}_{E_{A1(B1,C1)}}$ and $\hat{X}_{E_{A2(B2,C2)}}$ are homodyned, and can be obtained from
the reduced covariance matrix $\mathbf{V}_{A_1(B_3,C_3)|X_{E_{A1(B1, C1)}},X_{E_{A2(B2, C2)}}}$.
It is the maximal mutual information between Eve's measurement data and Alice's (Bob's, Charlie's) measurement data.
Because Eve can decrease $V(\hat{X}_{A1})$ most by homodyning on $\hat{X}_{E_{A1}}$  and
$\hat{X}_{E_{A2}}$ for reverse reconciliation, and reduce $V(\hat{X}_{B_3(C_3)})$ most by homodyning on $\hat{X}_{E_{B1(C1)}}$ and $\hat{X}_{E_{B2(C2)}}$ for direct reconciliation.

The first terms at the righthand sides of Eq. (\ref{eq5}) and Eq. (\ref{eq10}) are
\begin{equation}\label{eq17}
I(P_{A_1},P_{B_3}:P_{C_3})=\frac{1}{2}\log_2{\frac{V(\hat{P}_{C_3})}{V(\hat{P}_{C_3}|P_{A_1},P_{B_3})}}.
\end{equation}

The second term in Eq. (\ref{eq5}) is
\begin{equation}
 I(P_{C_3}:P_{E_{C1}},P_{E_{C2}})=\frac{1}{2}\log_2{\frac{V(\hat{P}_{C_3})}{V(\hat{P}_{C_3}|P_{E_{C1}},P_{E_{C2}})}},
\end{equation}
which is the maximal mutual information between Eve's measurement data and Charlie's measurement data with reverse reconciliation.
The second term in Eq. (\ref{eq10}) is
\begin{equation}
\begin{aligned}
 &I(P_{A_1}:P_{E_{A1}},P_{E_{A2}})+I(P_{B_3}:P_{E_{B1}},P_{E_{B2}})\\
 =&\frac{1}{2}\log_2{\frac{V(\hat{P}_{A_1})}{V(\hat{P}_{A_1}|P_{E_{A1}},P_{E_{A2}})}}+
 \frac{1}{2}\log_2{\frac{V(\hat{P}_{B_3})}{V(\hat{P}_{B_3}|P_{E_{B1}},P_{E_{B2}})}},
\end{aligned}
\end{equation}
which gives the maximal mutual information between Eve's measurement data and Alice's and Bob's measurement data with direct reconciliation.

\subsection{Coherent attack}
\begin{figure}
  \centering
  \includegraphics[width=0.4\textwidth]{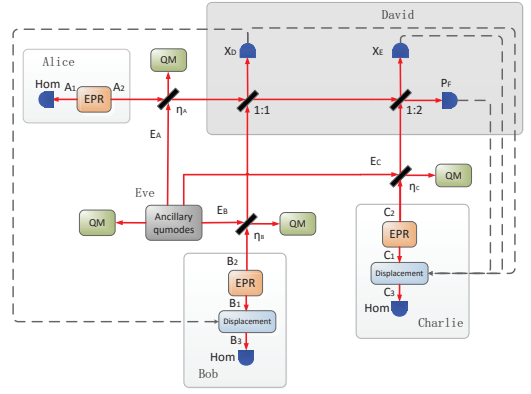}\\
  \caption{(Color online) EB scheme against a coherent attack}\label{fig8}
\end{figure}

In previous subsection, we analyze the security of our protocol under individual entangling cloner attacks. But this is not sufficient to show its unconditional security. In this subsection, we investigate the security of our protocol against a more general attack, which is a coherent attack within each time when Alice, Bob and Charlie respectively send one qumode to David. Note that a general coherent attack can be simplified to the coherent attack in Fig.~\ref{fig8} under the assumption that Alice, Bob and Charlie's input states are respectively permutationally symmetric~\cite{PhysRevLett.102.110504}.

Fig. \ref{fig8} shows a coherent attack against our protocol. Eve takes three qumodes out of his ancillary qumodes, which is globally a pure Gaussian state, and inject them into three channels through beam splitters, respectively. The output states coming out of the beam splitters and the remaining ancillary qumodes are all stored in Eve's quantum memory. After monitoring all the data in public channels, Eve implements an optimal measurement on these qumodes in the quantum memory to obtain maximal information.

We use the Holevo bound to quantify the amount of information Eve can obtain. The secret key rate in QCC protocol becomes
\begin{equation}
\begin{aligned}
K_{AB}^{RR}&=\beta I(X_{A_1}:X_{B_3})- H(\rho_{Eve}:X_{A_1}), \textrm{ and}\\
K_{AC}^{RR}&=\beta I(X_{A_1}:X_{C_3})- H(\rho_{Eve}:X_{A_1}),
\end{aligned}
\end{equation}
where $I(X_{A_1}:X_{B_3(C_3)})$ has been given in Eq. (\ref{eq11}), and $H(\rho_{Eve}:X_{A_1})=S(\rho_{Eve})-S(\rho_{Eve|X_{A_1}})$ denotes the Holevo information between Eve's quantum state and Alice's measurement data $X_{A_1}$. Here we use $S(\rho)$ to denote the von Neumann entropy of the quantum state $\rho$.
Since Eve can purify the whole state $\rho_{A_1,B_3,C_3,Eve}$, we have
\begin{equation}
H(\rho_{Eve}:X_{A_1}) =S(\rho_{A_1,B_3,C_3})-S(\rho_{B_3,C_3|X_{A_1}}),
\end{equation}
For $S(\rho_{A_1,B_3,C_3})$, we can calculate it from a function of the symplectic eigenvalues $\nu_1, \nu_2$ and $\nu_3$ of the covariance matrix $\mathbf{V}_{A_1,B_3,C_3}$.
\begin{equation}
  S(\rho_{A_1,B_3,C_3})=h(\nu_1)+h(\nu_2)+h(\nu_3),
\end{equation}
where $h(x):=\frac{x+1}{2}\log_2\frac{x+1}{2}-\frac{x-1}{2}\log_2\frac{x-1}{2}.$
For $S(\rho_{B_3,C_3|X_{A_1}})$, we have
\begin{equation}
  S(\rho_{B_3,C_3|X_{A_1}})=h(\nu_4)+h(\nu_5),
\end{equation}
where $\nu_4$ and $\nu_5$ are the symplectic eigenvalues of the covariance matrix $\mathbf{V}_{B_3,C_3|X_{A_1}}$.

The secret key rate for QSS scheme is
\begin{equation}
  K_{QSS}^{RR}=\beta I(P_{A_1},P_{B_3}:P_{C_3})- H(\rho_{Eve}:P_{C_3}),
\end{equation}
where $I(P_{A_1},P_{B_3}:P_{C_3})$ has been given in Eq. (\ref{eq17}), and
\begin{equation}
\begin{aligned}
  H(\rho_{Eve}:P_{C_3})=&S(\rho_{Eve})-S(\rho_{Eve|P_{C_3}})\\
  =&S(\rho_{A_1,B_3,C_3})-S(\rho_{A_1,B_3|P_{C_3}}).
\end{aligned}
\end{equation}
$S(\rho_{A_1,B_3|P_{C_3}})$ can be calculated from $h(\nu_6)+h(\nu_7)$, where $\nu_6$ and $\nu_7$ are the symplectic eigenvalues of the covariance matrix $\mathbf{V}_{A_1,B_3|P_{C_3}}$.

Both the secret key rates for QCC and QSS are functions of the elements of the covariance matrix $\mathbf{V}_{A_1,B_3,C_3}$ and its reduced covariance matrix under partial homodyne measurement.
In the following, we show how to obtain the covariance matrix $\mathbf{V}_{A_1,B_3,C_3}$.

At the beginning, the whole system is the tensor product of Alice's, Bob's and Charlie's TMSS's and Eve's globally pure Gaussian state. Generally, up to local Gaussian operation, the covariance matrix of Eve's reduced state $\rho_{E_A, E_B, E_C}$ in Fig.~(\ref{fig8}) can be given by
\begin{equation} \label{eq2}
\mathbf{V_{E_A, E_B, E_C}}=\left(
  \begin{array}{ccc}
    \mathbf{V_A} & \mathbf{G_1} & \mathbf{G_2} \\
    \mathbf{G_1} & \mathbf{V_B} & \mathbf{G_3} \\
    \mathbf{G_2} & \mathbf{G_3} & \mathbf{V_C} \\
  \end{array}
\right),
\end{equation}
where
\begin{equation}
\begin{aligned}
  &\mathbf{V_{E_A}}= V_{E_A}\mathbf{I}, \, \mathbf{V_{E_B}}=V_{E_B}\mathbf{I}, \, \mathbf{V_{E_C}}=V_{E_C}\mathbf{I},\\
  &\mathbf{G_1}=\left(
                 \begin{array}{cc}
                   g_1 & 0 \\
                   0 & g'_1 \\
                 \end{array}
               \right), \,
  \mathbf{G_2}=\left(
                 \begin{array}{cc}
                   g_2 & 0 \\
                   0 & g'_2 \\
                 \end{array}
               \right),  \\
               &\textrm{and }
  \mathbf{G_3}=\left(
                 \begin{array}{cc}
                   g_3 & 0 \\
                   0 & g'_3 \\
                 \end{array}
               \right).
  \end{aligned}
\end{equation}
 $V_{E_A}, V_{E_B}$ and $V_{E_C}$ are the variances of the thermal noise Eve injects into each channel. $g_1, g_2$ and $g_3$ represent the correlations between the noises Eve adds into the three channels.
Then the covariance matrix of the whole system can be written as
\begin{equation}
\mathbf{V}_{A, B, C, Eve}=
\bigoplus_{k=1}^3\mathbf{V'}\bigoplus\mathbf{V}_{E_A, E_B, E_C},
\end{equation}
where
\begin{equation}
\mathbf{V'}=\left(
  \begin{array}{cc}
    V\mathbf{I} & \sqrt{V^2-1}\mathbf{Z}  \\
    \sqrt{V^2-1}\mathbf{Z} & V\mathbf{I} \end{array}
    \right).
\end{equation}

Permutate the modes in the covariance matrix $\mathbf{V}_{A, B, C, Eve}$ to make the order of the modes becomes $A,E_A,B,E_B,C,E_C$.
Applying the conjugate unitary operation on the covariance matrix $\mathbf{V}_{A,E_A,B,E_B,C,E_C}$, we obtain the covariance matrix of the whole state including the modes $A_1, B_1, C_1$ and Eve's modes, that is
\begin{equation} \label{eq16}
  \mathbf{U}_{David}\mathbf{U}_{Eve} \mathbf{V}_{A,E_A,B,E_B,C,E_C}\mathbf{U}_{Eve}^T\mathbf{U}_{David}^T.
\end{equation}
 Here the matrices $\mathbf{U}_{David}$ and $\mathbf{U}_{Eve}$ are different from those given in Eqs. (\ref{eq12}),(\ref{eq13}),(\ref{eq14}) and (\ref{eq15}). We must delete the seventh and eighth rows and columns of the matrices $\mathbf{BS}_{A(B,C)}$ in Eq. (\ref{eq13}) and $\mathbf{W}_{1(2,3,4)}$ in Eq. (\ref{eq15}), to make the dimensions of the matrices $\mathbf{U}_{David}$ and $\mathbf{U}_{Eve}$ match $\mathbf{V}_{A,E_A,B,E_B,C,E_C}$.
Since we want to obtain the covariance matrix $\mathbf{V}_{A_1,B_3,C_3}$, we delete the rows and the columns corresponding to Eve's modes in the covariance matrix given by Eq. (\ref{eq16}).
Then we permutate the modes in the order $A_1, B_1, C_1, D, E, F$, obtaining the covariance matrix $\mathbf{V}_{A_1,B_1,C_1,D,E,F}$. Using the formula of reduced covariance matrix following partial homodyne measurement as shown in Eq. (\ref{eq8}), we get the covariance matrix $\mathbf{V}_{A_1,B_1,C_1|X_D,X_E,P_F}$. Since displacement operations don't change the covariance matrix, $\mathbf{V}_{A_1,B_1,C_1|X_D,X_E,P_F}$  is the covariance matrix $\mathbf{V}_{A_1,B_3,C_3|X_D,X_E,P_F}$.

\section{Simulation results}
\label{sec3}
In this section, we simulate both QCC and QSS schemes against two kinds of attacks according with the state-of-art technology. The simulation results show that under independent entangling cloner attacks, the maximal transmission distances can be significantly enlarged in the case of unbalanced $L_A, L_B$ and $L_C$. But under coherent attacks, the maximal transmission distances are markedly reduced. By comparing different entangled attacks, we finally find the optimal coherent attacks in QCC and QSS.

\subsection{Simulation for independent entangling cloner attack}

We can replace the transmittance of the beam splitter in Fig.~\ref{fig3}, with a realistic transmission distance  in experiment by using $\eta_{A\,(B,C)}=10^{-\alpha\frac{L_{A(B,C)}}{10}}$, where $L_{A(B,C)}$ is the transmission distance from Alice (Bob, Charlie) to David, and $\alpha$ denotes the coefficient of loss in optical fibers. In the following simulation, we set the coefficient of loss $\alpha=0.2dB/km$. 

Besides the transmission distances, the secret key rate also depends on the variance of Alice's (Bob's, Charlie's) initial TMSS's, $V$, the variance of Eve's TMSS's, $V_E$  and the reconciliation efficiency $\beta$. According to the current state-of-art experimental technology, we set $V=10$ and $\beta=0.95$ in the following simulation. Since larger $V_E$ indicates higher noise in the channel and lower uncertainty for Eve's estimation, we set $V_E=1$ for pure loss case, $V_E=2$ for weak entangling cloner attack and $V_E=5$ for strong entangling cloner attack in our simulation.

\begin{figure}[htbp]
  \centering
  \psfrag{t}[][]{$L_{B(C)}/km$}
  \psfrag{s}[][]{$L_{A}/km$}
  \includegraphics[width=0.4\textwidth]{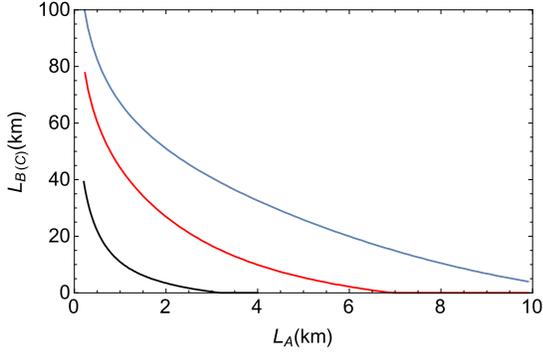}\\
  \caption{(Color online) The maximal transmission distances satisfying the condition $K_{AB(AC)}^{RR}>10^{-3}$. The blue curve is for the case $V_E=1$; the red curve is for the case $V_E=2$; the black is for the case $V_E=5$.}\label{fig4}
\end{figure}

In QCC, we assume that the transmission distances from Bob and Charlie to David are equal, i.e., $L_B=L_C$, while the transmission distance from Alice to David, $L_A$, is different from $L_B$ and $L_C$. With reverse reconciliation, when Alice is close to David, the transmittance $\eta_A=10^{-0.2\frac{L_{A}}{10}}$ approaches $1$, so that little information can be obtained by Eve from Alice's measurement data. So the secure transmission distance from Bob and Charlie to David can be significantly increased. Fig.~\ref{fig4} shows the maximal transmission distances of $L_{A}$ and $L_{B(C)}$ satisfying $K_{AB(AC)}^{RR}>10^{-3}$. With direct reconciliation, the situation is opposite. To attain a high secret key rate, Bob and Charlie must be close to David, while Alice can be far away from David.

\begin{figure}[htbp]
  \centering
  \psfrag{s}[][]{$L_{C}/km$}
  \psfrag{t}[][]{$L_{A(B)}/km$}
  \includegraphics[width=0.4\textwidth]{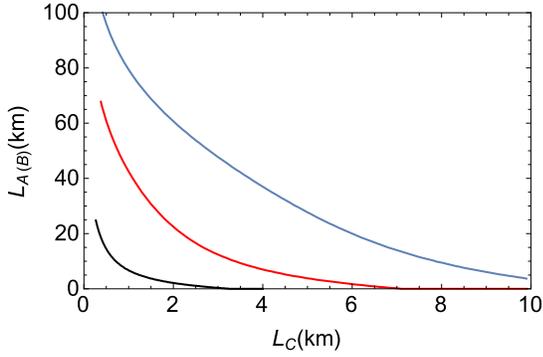}\\
  \caption{(Color online) The maximal transmission distances satisfying the condition $K_{QSS}^{RR}>10^{-3}$. The blue curve is for the case $V_E=1$; the red curve is for the case $V_E=2$; the black is for the case $V_E=5$.}\label{fig5}
\end{figure}

In QSS, we consider the case that $L_A=L_B$, but $L_C$ is different from $L_A$ and $L_B$. With reverse reconciliation, when $L_{C}$ approaches zero, the transmittance $\eta_C=10^{-0.2\frac{L_{C}}{10}}$ gets close to $1$, so that Eve can obtain little amount of information from Charlie's measurement data. Hence, both the secure transmission distances $L_{A}$ and $L_{B}$ are greatly enlarged as shown in Fig.~\ref{fig5}. With direct reconciliation, to keep the secret key rate high, Alice can be far from David, but both Alice and Bob must be close to David.

The simulation results in both Fig.~\ref{fig4} and Fig.~\ref{fig5} show that
imbalanced transmission distances of the three channels lead to further total maximal transmission distances when Eve implements entangling cloner attack.

\subsection{Simulation for coherent attack}
To guarantee the matrix $\mathbf{V_{E_A, E_B, E_C}}$ in Eq. (\ref{eq2}) a valid covariance matrix, for any thermal noise $V_{E_A}, V_{E_B}, V_{E_C}\ge 1$, $g_1, g_2$ and $g_3$ must satisfy the bona-fide condition~\cite{1751-8121-40-28-S01}, that is $\nu^2_{-}\ge1$, where $\nu_{-}$ is the smallest symplectic eigenvalue of the matrix $\mathbf{V_{E_A, E_B, E_C}}$. The symplectic eigenvalue spectrum of the matrix $\mathbf{V_{E_A, E_B, E_C}}$ equals to the eigenvalue spectrum of the matrix $|i\mathbf{\Omega} \mathbf{V_{E_A, E_B, E_C}}|$, where
\begin{equation}
  \mathbf{\Omega}=\bigoplus_{k=1}^3 \left(
                                   \begin{array}{cc}
                                     0 & 1 \\
                                     -1 & 0 \\
                                   \end{array}
                                 \right).
\end{equation}

The secret key rates depend on the transmission distances, the variance of Alice (Bob, Charlie)'s TMSS's, $V$, the thermal noise Eve injects in each channel, denoted by $V_{E_A}, V_{E_B}$ and $V_{E_C}$, the correlations between the noises in the three channels, represented by $g_1, g_2$ and $g_3$, and the reconciliation efficiency $\beta$. Here we set $V=10$ and $\beta=0.95$, same as above. 

To minimize the secret key rate in Eq. (\ref{eq1}), Eve only needs to concentrate on attacking the communication either between Alice and Bob, or between Alice and Charlie using the optimal ``negative EPR attack", which has been defined in refs.~\cite{PhysRevA.91.022320,pirandola2015high}.   

Another case, which has not been investigated before, is that Eve intends to reduce the secret key rates $K_{AB}$ and $K_{AC}$ simultaneously, such that Alice can securely communicate neither with Bob, nor with Charlie. To do this, one way for Eve is to apply symmetric attack in Bob's and Charlie's channels, which means that interchanging $\rho_{E_B}$ and $\rho_{E_C}$ leaves Eve's attack invariant, i.e., $V_{E_B}=V_{E_C}$ and $g_1=g_2$. 

\begin{figure}[htbp]
  \centering
  \psfrag{x}[][]{$g_1 (g_2)$}
  \psfrag{y}[][]{$g_3$}
  \includegraphics[width=0.4\textwidth]{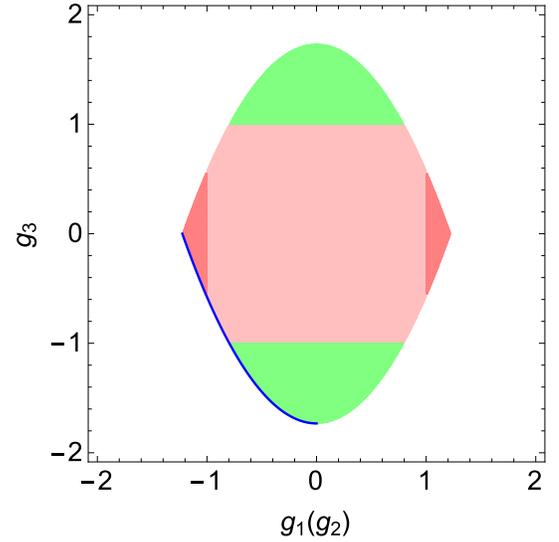}\\
  \caption{(Color online) The accessible values of $g_1 (g_2)$ and $g_3$ satisfying the bona-fide condition when $V_{E_A}=V_{E_B}=V_{E_C}=2$. The red region shows the values of $g_1 (g_2)$ and $g_3$, with which $\rho_{E_A}$ is bipartitely entangled with $\rho_{E_B}$ and $\rho_{E_C}$ respectively. The green region shows the values of $g_1 (g_2)$ and $g_3$, with which $\rho_{E_B}$ and $\rho_{E_C}$ are entangled.
  }\label{fig10}
\end{figure}

We find that the bona-fide condition in this symmetric case becomes 
\begin{align}\notag
&4g_1^2+g_3^2-V_{E_A}^2-V_{E_C}^2+\\
&\sqrt{8g_1^2[g_3^2-(V_{E_A}-V_{E_C})^2]+(g_3^2+V_{E_A}^2-V_{E_C}^2)^2}\le -2. \label{eq17}
\end{align}
A numerical example of the accessible values of $g_1 (g_2)$ and $g_3$ satisfying the bona-fide condition is shown by the colored region in Fig. \ref{fig10}.   We divide this colored region into three subregions by checking the separability of each two modes in their reduced states using the positive partial transpose (PPT) criterion~\cite{PhysRevLett.84.2726}.
In the red region, $\rho_{E_A}$ is entangled with $\rho_{E_B}$ and $\rho_{E_C}$ respectively. In the green region, $\rho_{E_B}$ and $\rho_{E_C}$ are entangled. In the pink region, $\rho_{E_A}, \rho_{E_B}$ and $\rho_{E_C}$ are pairwise separable. 

 In the left red region, the fluctuations of $\hat{X}_{E_A}-\hat{X}_{E_B}$ and $\hat{X}_{E_A}-\hat{X}_{E_C}$ are amplified. Injecting this kind of noise results in the increase of the fluctuations of $\hat{X}_{A_1}-\hat{X}_{B_3}$ and $\hat{X}_{A_1}-\hat{X}_{C_3}$. In the bottom green region, the fluctuation of $\hat{X}_{E_B}-\hat{X}_{E_C}$ is amplified. Injecting this kind of noise makes the fluctuation of $\hat{X}_{B_3}-\hat{X}_{C_3}$ increase. For both cases, the secret key rate is declined. Referring to the discussion in ref.~\cite{pirandola2015high,PhysRevA.91.022320}, the entanglement corresponding to the left red region and the bottom green region is ``bad" entanglement.
 Whereas, the entanglement corresponding to the right red region and the top green region helps increasing the key rate, which we call ``good" entanglement. 
 To minimize the secret key rate, Eve needs to maximize the ``bad" correlation between the noises in each channel. Thus, the optimal attack must lie in the blue contour curve in Fig.~\ref{fig10}. 
 
\begin{figure}[htbp]
  \centering
  \psfrag{t}[][]{$L_{B(C)}/km$}
  \psfrag{s}[][]{$L_{A}/km$}
  \includegraphics[width=0.4\textwidth]{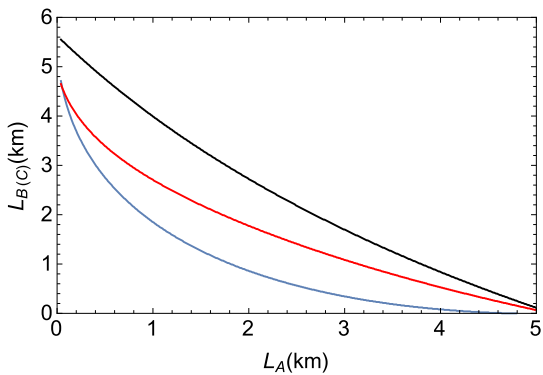}\\
  \caption{(Color online) The maximal transmission distances satisfying $K_{AB(AC)}^{RR}>0$ when $V_{E_A}=V_{E_B}=V_{E_C}=2$.  The blue curve is the attack with minimal $g_1(g_2)$ and vanishing $g_3$. The red curve is the attack with minimal $g_3$ and vanishing $g_1(g_2)$. The black curve is the attack with $g_1=g_2=g_3=0$.}\label{fig11}
\end{figure}

We first compare two maximally entangled attacks with the independent attack.  The first entangled attack corresponds to the left-most point in the red region, where $g_1(g_2)$ is minimized and $g_3=0$. The second entangled attack corresponds to the down-most point in the green region,  where $g_3$ is minimized and $g_1=g_2=0$. The independent attack corresponds to the origin given by $g_1=g_2=g_3=0$. The simulation results of these three attacks are given in Fig.~\ref{fig11}. It  indicates that entangled attacks perform better than independent attack and the attack with minimal $g_1(g_2)$ and zero $g_3$ is stronger than the other two attacks.  

\begin{figure}[htbp]
  \centering
  \psfrag{t}[][]{$L_{B(C)}/km$}
  \psfrag{s}[][]{$L_{A}/km$}
  \includegraphics[width=0.4\textwidth]{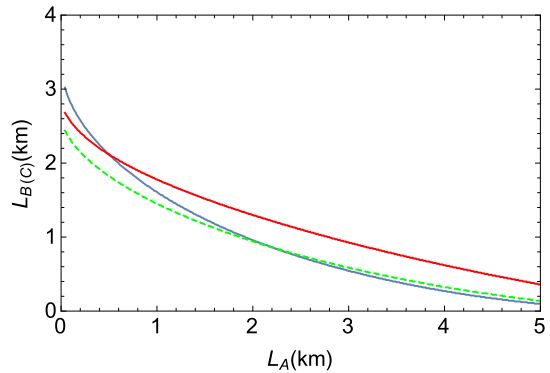}\\
  \caption{(Color online) The maximal transmission distances satisfying $K_{AB(AC)}^{RR}>0$ when $V_{E_A}=1.5$ and $V_{E_B}=V_{E_C}=3$. The blue curve is the attack with $g_1(g_2)$ minimized and $g_3$ vanishing; the red curve is the attack with $g_1(g_2)$ vanishing and $g_3$ minimized; the dashed green curve is the attack, where $g_1(g_2)$ equals $2/3$ times of its minimal value and $g_3$ saturates the bona-fide condition in Eq.~(\ref{eq17}).}\label{fig13}
\end{figure}

But the attack with minimal $g_1(g_2)$ and zero $g_3$ is not a general optimal attack. For instance, in the case shown by Fig.~\ref{fig13}, when $L_A$ is short, the attack corresponding to the dashed green curve performs better than the attack with minimal $g_1(g_2)$ and vanishing $g_3$, shown by the blue curve.  
We cannot give a universal optimal attack strategy of $g_1(g_2)$ and $g_3$ for all the cases. The values of $g_1(g_2)$ and $g_3$ to achieve optimal attack depend on the transmission distance of each channel and the thermal noise in each channel.

\begin{figure}[htbp]
  \centering
  \psfrag{s}[][]{$V_{E_{A}}/km$}
  \psfrag{t}[][]{$V_{E_{B(C)}}/km$}
  \includegraphics[width=0.4\textwidth]{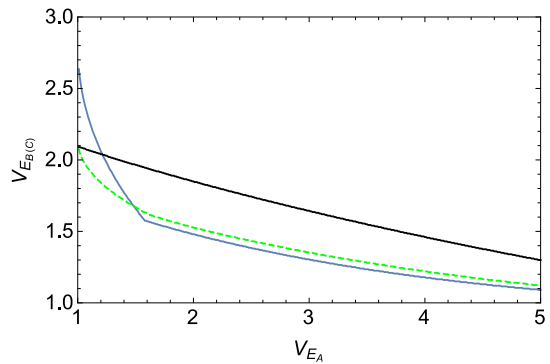}\\
  \caption{(Color online) The maximal tolerable thermal noises satisfying $K_{AB(AC)}^{RR}>0$ when $L_A=1\,km$ and $L_B=L_C=3\,km$. The blue curve is the attack with $g_1(g_2)$ minimized and $g_3$ vanishing; the black curve is the attack with $g_1(g_2)$ vanishing and $g_3$ minimized; the dashed green curve is the attack, where $g_1(g_2)$ equals half of its minimal value and $g_3$ saturates the bona-fide condition in Eq.~(\ref{eq17}).}\label{fig15}
\end{figure}

In Fig.~\ref{fig11} and Fig.~\ref{fig13}, we fix the thermal noise in each channel to see the maximal transmission distances. Now we fix the transmission distances to look at the maximal tolerable thermal noise in each channel. Fig.~\ref{fig15} gives the simulation when $L_A=1\,km$ and $L_B=L_C=3\,km$. The simulation result indicates that the maximal tolerable $V_{E_A}$ and $V_{E_B(C)}$ is markedly asymmetric. 

For QSS, we consider the symmetric case where the transmission distances of Alice's and Bob's channels are equal, $L_A=L_B$, and that the attacks in Alice's and Bob's channels are the same, $V_{E_A}=V_{E_B}$ and $g_2=g_3$. It implies that after Charlie distributes the secret to Alice and Bob, both of them can obtain the same amount of information about the secret.

\begin{figure}[htbp]
  \centering
  \psfrag{x}[][]{$g_2 (g_3)$}
  \psfrag{y}[][]{$g_1$}
  \includegraphics[width=0.4\textwidth]{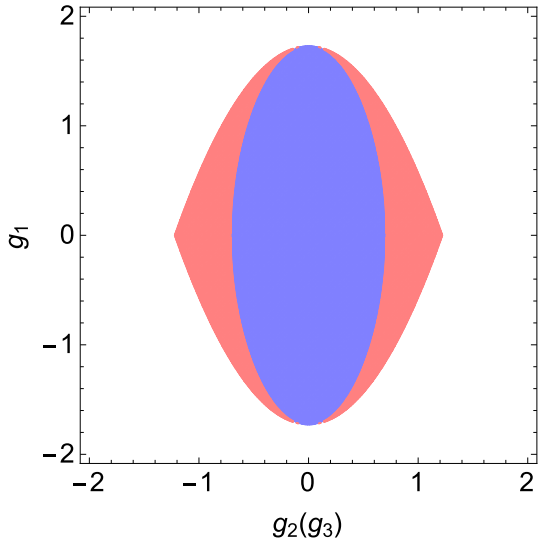}\\
  \caption{(Color online) The accessible values of $g_1$ and $g_2 (g_3)$ satisfying the bona-fide condition when $V_{E_A}=V_{E_B}=V_{E_C}=2$. The red region shows the values of $g_1$ and $g_2 (g_3)$, with which $\rho_{E_C}$ is entangled with $\rho_{E_A,E_B}$. The blue region shows the values of $g_1$ and $g_2 (g_3)$, with which $\rho_{E_C}$ and $\rho_{E_A, E_B}$ are separable.
  }\label{fig14}
\end{figure}

\begin{figure}[htbp]
  \centering
  \psfrag{t}[][]{$L_{C}/km$}
  \psfrag{s}[][]{$L_{A(B)}/km$}
  \includegraphics[width=0.4\textwidth]{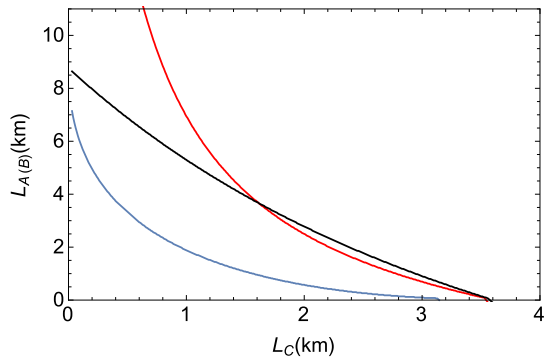}\\
  \caption{(Color online) The maximal transmission distances satisfying $K_{QSS}^{RR}>0$ when $V_{E_A}=V_{E_B}=2$ and $V_{E_C}=3$. The blue curve is the attack with minimal $g_2(g_3)$ and vanishing $g_1$; the red curve is the attack with vanishing $g_2(g_3)$ and minimal $g_1$; the black curve represents independent attack with $g_1=g_2=g_3=0$.}\label{fig12}
\end{figure}

A numerical example of the accessible values of $g_1$ and $g_2(g_3)$ is shown in Fig.~\ref{fig14}. By using the PPT criterion, we divide the colored region further into two subregions. $\rho_{E_C}$ and $\rho_{E_A, E_B}$ are separable in the blue region and entangled in the red region. For the entangled states in the left red region, the fluctuations of $\hat{P}_{E_A}+\hat{P}_{E_C}$ and $\hat{P}_{E_B}+\hat{P}_{E_C}$ are simultaneously amplified, which makes the fluctuation of $\hat{P}_{A_1}+\hat{P}_{B_3}+\hat{P}_{C_3}$ increase. So it is ``bad" entanglement, helping Eve to decrease the secret key rate. Whereas the entanglement in the right red region is ``good" entanglement, helping to increase the secret key rate. For any $V_{E_C}$ and $V_{E_A}(V_{E_B})$, the optimal attack corresponds to the attack with minimal $g_2(g_3)$ and vanishing $g_1$, which indicates that $\rho_{E_C}$ and $\rho_{E_A,E_B}$ are maximally entangled.

We compare the optimal attack with other two attacks. One is the independent attack given by $g_1=g_2=g_3=0$. The other is the attack with minimal $g_1$ and vanishing $g_2(g_3)$, in which $\rho_{E_A}$ and $\rho_{E_B}$ form an EPR pair and $\rho_{E_C}$ is independent.  
Fig.~\ref{fig12} shows the maximal transmission distances of $L_{A(B)}$ and $L_C$ under these three attacks when $V_{E_A}=V_{E_B}=2$ and $V_{E_C}=3$. It's easy to see that the attack with minimal $g_2(g_3)$ performs better than the other two attacks.

\section{Conclusion}
\label{sec4}
This paper investigates CV MDI multipartite quantum communication, where detector side attacks are removed from the side of each party participating the quantum communication. Our protocol can implement both QCC and QSS. The security against entangling cloner attack and coherent attack is analyzed respectively. Under entangling cloner attack, the maximal transmission distances can be significantly enlarged in the case of unbalanced distribution. Compared with entangling cloner attack, coherent attack reduces the maximally transmission distances markedly. Finally, we study the optimal coherent attacks in QCC and QSS respectively.

\begin{acknowledgements}
This work is supported by the National Natural Science Foundation of China (Grant Nos. 61475099, 61102053, 61379153?61378012), Program of State Key Laboratory of Quantum Optics and Quantum Optics Devices (No: KF201405), Open Fund of IPOC (BUPT) (No: IPOC2015B004).
\end{acknowledgements}

\bibliographystyle{apsrev4-1}
\bibliography{reference}
\end{document}